\documentclass[a4paper,11pt]{article}
\pdfoutput=1 % if your are submitting a pdflatex (i.e. if you have
\usepackage{jcappub} % for details on the use of the package, please
\usepackage[utf8]{inputenc}
\usepackage[T1]{fontenc}

\usepackage[mathscr]{eucal}
\usepackage[Symbolsmallscale]{upgreek}
\usepackage{xcolor}
\usepackage{slashed}
\usepackage{hyperref}
\usepackage{soul}
\usepackage[normalem]{ulem}

\usepackage{graphicx}
\usepackage{subfigure}
\usepackage{hyperref}
\usepackage{ulem}
\usepackage{float}
\usepackage{paracol}
\usepackage{placeins}
\usepackage{blindtext}

\usepackage{xcolor}
\usepackage[utf8]{inputenc}

\begin{document}

\title{Did we hear the sound of the Universe boiling? \\ \it{Analysis using the full
fluid velocity profiles and NANOGrav 15-year data}}

% \author{Tathagata Ghosh}
% \email[tathagataghosh@hri.res.in]{...}
% \affiliation{Harish-Chandra Research Institute, \\ A CI of Homi Bhabha National Institute, Chhatnag Road, Jhusi, Prayagraj 211019, India}
% \author{Anish Ghoshal}
% \email[anish.ghoshal@fuw.edu.pl]{...}
% \affiliation{Institute of Theoretical Physics, Faculty of Physics, University of Warsaw, \\ ul. Pasteura 5, 02-093 Warsaw, Poland}
% \author{Huai-Ke Guo}
% \email[guohuaike@ucas.ac.cn]{...}
% \affiliation{
% International Centre for Theoretical Physics Asia-Pacific, University of Chinese Academy of Sciences, 100190 Beijing, China
% }
% \author{Fazlollah Hajkarim}
% \email[fazlollah.hajkarim@ou.edu]{...}
% \affiliation{Homer L. Dodge Department of Physics and Astronomy, University of Oklahoma, Norman, OK 73019, USA}
% \author{Stephen F King}
% \email[king@soton.ac.uk]{...}
% \affiliation{School of Physics and Astronomy, University of Southampton,\\
% 	Southampton SO17 1BJ, United Kingdom}
% \author{Kuver Sinha}
% \email[kuver.sinha@ou.edu]{...}
% \affiliation{Homer L. Dodge Department of Physics and Astronomy, University of Oklahoma, Norman, OK 73019, USA}
% \author{Xin Wang}
% \email[Xin.Wang@soton.ac.uk]{...}
% \affiliation{School of Physics and Astronomy, University of Southampton,\\
% 	Southampton SO17 1BJ, United Kingdom}
% \author{Graham White}
% \email[g.a.white@soton.ac.uk]{...}
% \affiliation{School of Physics and Astronomy, University of Southampton,\\
% 	Southampton SO17 1BJ, United Kingdom}

\author[a]{Tathagata Ghosh,}
\author[b]{Anish Ghoshal,}
\author[c]{Huai-Ke Guo,}
\author[d]{Fazlollah Hajkarim,}
\author[e]{Stephen F King,}
\author[d]{Kuver Sinha,}
\author[e]{Xin Wang,}
\author[e]{Graham White,}

\affiliation[a]{Harish-Chandra Research Institute,  A CI of Homi Bhabha National Institute, Chhatnag Road, Jhusi, Prayagraj 211019, India}

\affiliation[b]{Institute of Theoretical Physics, Faculty of Physics, University of Warsaw, ul. Pasteura 5, 02-093 Warsaw, Poland}

\affiliation[c]{International Centre for Theoretical Physics Asia-Pacific, University of Chinese Academy of Sciences, 100190 Beijing, China}

\affiliation[d]{Homer L. Dodge Department of Physics and Astronomy, University of Oklahoma, Norman, OK 73019, USA}

\affiliation[e]{School of Physics and Astronomy, University of Southampton, Southampton SO17 1BJ, United Kingdom}

\emailAdd{tathagataghosh@hri.res.in}
\emailAdd{anish.ghoshal@fuw.edu.pl}
\emailAdd{guohuaike@ucas.ac.cn}
\emailAdd{fazlollah.hajkarim@ou.edu }
\emailAdd{king@soton.ac.uk}
\emailAdd{kuver.sinha@ou.edu}
\emailAdd{Xin.Wang@soton.ac.uk}
\emailAdd{g.a.white@soton.ac.uk}

% \date{\today}

\abstract{
In this paper, we analyse sound waves arising from a cosmic phase transition where the full velocity profile is taken into account as an explanation for the gravitational wave spectrum observed by multiple pulsar timing array groups. Unlike the broken power law used in the literature, in this scenario the power law after the peak depends on the macroscopic properties of the phase transition, allowing for a better fit with pulsar timing array (PTA) data. We compare the best fit with that obtained using the usual broken power law and, unsurprisingly, find a better fit with the gravitational wave (GW) spectrum that utilizes the full velocity profile. Even more importantly, the thermal parameters that produce the best fit are quite different. We then discuss models that can produce the best-fit point and complementary probes using CMB experiments and searches for light particles in DUNE, IceCUBE-Gen2, neutrinoless double $\beta-$decay, and forward physics facilities (FPF) at the LHC like FASER$\nu$, etc.
}%\end{abstract}

\maketitle

\section{Introduction}

It has been known for some time that Pulsar Timing Array (PTA) experiments can be used to detect gravitational waves (GWs)~\cite{Sazhin1978, Detweiler:1979wn, Foster1990}. This is possible by studying the timing distortions of successive light pulses emitted by millisecond pulsars, which are extremely stable clocks. The PTAs search for spatially correlated fluctuations in the pulse arrival time measurements of such pulsars, due to GWs perturbing the space-time metric along the line of sight to each pulsar. For GWs, the timing distortions should exhibit the angular dependence expected for an isotropic background of spin two GWs which enables them to be distinguished from either spin-zero or spin-one waves, and other effects, according to the work of Hellings and Down~\cite{Hellings:1983fr}. 

Recently, several PTA projects have reported the discovery of a stochastic gravitational wave background (SGWB). In particular, the North American Nanohertz Observatory for Gravitational Waves (NANOGrav)~\cite{NANOGrav:2023gor}, the European PTA~\cite{Antoniadis:2023ott}, the Parkes PTA~\cite{Reardon:2023gzh} and the Chinese CPTA~\cite{Xu2023} have all released results which seem to be consistent with a Hellings-Downs pattern in the angular correlations which is characteristic of the SGWB. In particular, the largest statistical evidence for SGWB is seen in the NANOGrav 15-year data (NANOGrav15)~\cite{NANOGrav:2023gor}. This is the first discovery of GWs in the frequency around $10^{-8}$ Hz, and wavelengths around 10 light years.
The most obvious origin of such an SGWB is due to the merging supermassive black hole binaries (SMBHBs) resulting from the collision of two galaxies, each with an SMBH with masses in the range $10^{8-9}$ solar masses at its centre~\cite{Sesana:2004sp, Burke-Spolaor:2018bvk}. The expected amplitude has an order of magnitude uncertainty depending on the density, redshift, and other properties of SMBH sources. Indeed, there may be millions of such sources contributing to the SGWB.

However, the current data does not allow individual SMBH binary sources to be identified, so it is unclear if the observed SGWB has an astrophysical or cosmological origin~\cite{NANOGrav:2023hvm, Antoniadis:2023xlr}. For example, the cosmological origin of SGWB could be due to first-order phase transitions~\cite{Winicour1973, Hogan:1986qda, Athron:2023xlk, Caprini:2010xv,  NANOGrav:2021flc,  Xue:2021gyq, DiBari:2021dri, Madge:2023cak}, cosmic strings~\cite{Siemens:2006yp, Ellis:2020ena, King:2020hyd, Buchmuller:2020lbh, Blasi:2020mfx, Bian:2022tju, Fu:2023nrn}, domain walls~\cite{Ferreira:2022zzo,  An:2023idh, Dunsky:2021tih}, or scalar-induced gravitational waves (SIGWs) generated from primordial fluctuations~\cite{Vaskonen:2020lbd, DeLuca:2020agl, Inomata:2020xad, Sugiyama:2020roc, Zhou:2020kkf, Ghoshal:2023sfa, Chen:2019xse}. Such possibilities represent new physics beyond the standard model (BSM) and it would be interesting to know how such alternative scenarios could be distinguished.

One characteristic feature is the shape of the spectrum in the recent data, which, unlike the previous results, seems to be blue-tilted~\cite{NANOGrav:2023gor, NANOGrav:2023hvm}. The analysis of the NANOGrav 12.5-year data release suggested a nearly flat GW spectrum as a function of frequency ($f$), 
$\Omega_{GW} \propto f^{(-1.5,0.5)}$ at one sigma, in a narrow range of frequencies around 5.5 nHz~\cite{NANOGrav:2020bcs}. By contrast, the recent 15-year data release finds a steeper slope, 
$\Omega_{GW} \propto f^{(1.3,2.4)}$ at one sigma~\cite{NANOGrav:2023gor}. The naive scaling predicted for GW from SMBH binaries is disfavoured by the latest NANOGrav data, although environmental and statistical effects can lead to different predictions~\cite{NANOGrav:2023hvm, Agazie2023}.

Motivated by the above considerations, new analyses are necessary to explore which SGWB formation mechanisms can lead to the generation of a signal consistent with these updated observations. Indeed, following the recent announcements, several papers have appeared which address some of these issues \cite{King:2023cgv, Megias2023, Han2023, Guo2023, Yang2023, Kitajima2023, Bai2023, Zu2023, Kitajima2023a, Vagnozzi2023, Lambiase2023, Ellis2023, Li2023, Franciolini2023, Shen2023, Ellis2023a, Franciolini2023a, Wang2023, Ghoshal2023, Fujikura2023, Athron:2023mer, Kitajima:2023vre, Lazarides:2023ksx, Yang:2023qlf, Addazi:2023jvg, Broadhurst:2023tus, Cai:2023dls, Inomata:2023zup, Depta:2023qst, Depta:2023qst, Eichhorn:2023gat, Huang:2023chx, Gouttenoire:2023ftk,Blasi:2023sej}.

In this paper, we consider the sound waves arising from a cosmic phase transition where the full velocity profile is taken into account. We compare the best fit with that obtained using the usual broken power law and find a better fit to NANOGrav data using the full velocity profile. We first explain how to obtain this result before discussing some models that can produce such thermal parameters. Finally, we discuss complementary probes of hidden sectors.

\section{PTA data and the sound shell model}

Multiple PTA collaborations observed compelling evidence\footnote{The goal of our paper is to assume that the evidence is robust and then ask the following theoretical questions: to what extent can the data be explained  by incorporating the effect of SMBH binaries; and can one consider a new physics source, especially sound waves of a first order phase transition? The extensive BSM studies performed by the  NANAGrav collaboration assume a similar outlook \cite{NANOGrav:2023hvm}. } for a gravitational wave spectrum, with NANOGrav and EPTA giving the best fit for a power law spectrum parametrized as follows,
\begin{equation}
    \Omega = \frac{8 \pi ^4 f^5  \Phi (f)}{H_0^2 \Delta f}
\end{equation}
with
\begin{equation}\label{para1}
    \Phi = \frac{A^2}{12 \pi ^2 T_{\rm obs}} \left( \frac{f}{{\rm yr} ^{-1} }\right) ^{- \gamma } {\rm yr} ^3
\end{equation}
where $\Delta f =1/T_{\rm obs}$ and $H_0=h \times 100 \, {\rm km} \, {\rm s}^{-1} {\rm Mpc}^{-1}$ is the current value of the Hubble rate. The best fit values of the parameters $A$ and $\gamma$ in Eq.~\ref{para1}  are given by
\begin{eqnarray}
    \gamma &=& \left\{ \begin{array}{cc} 3.2 \pm 0.6 & {\rm NANOGrav} \\ 
    3.1 ^{+0.77} _{-0.68} & {\rm EPTA}
    \end{array} \right. \\
    A &=& \left\{ \begin{array}{cc} 6.4^{+4.2}_{-2.7} \times 10^{-15} & {\rm NANOGrav} \\ 10^{-14.13 \pm 0.12} & {\rm EPTA} \end{array} \right.
\end{eqnarray}

 While inspiralling SMBHBs provide the standard astrophysical explanation for the signal, a  first-order  phase transition (FOPT) at the $\mathcal{O}({\rm MeV})$ scale is an intriguing alternative. In this Section, we model the FOPT with the sound shell model \cite{Hindmarsh:2016lnk}, obtain the corresponding GW spectrum, and compare our results with the fit performed by the NANOGrav Collaboration. 

The GW spectrum from a FOPT is characterized by the following parameters: the nucleation temperature $T_n$, the strength of the FOPT $\alpha_n$, the average separation of bubbles $R_n$ which can be related to the bubble nucleation rate $\beta$,  and the bubble wall velocity  $v_w$.    The fit frequently appearing in the literature, and in particular in the recent analysis of the NANOGrav paper describes a single broken power-law of the form \cite{Guo:2020grp, Guo:2021qcq, Caldwell:2022qsj, Espinosa:2010hh, Giese:2020rtr, Hindmarsh:2016lnk, Hindmarsh:2019phv, Hindmarsh:2017gnf}
\begin{eqnarray}
h^2 \Omega_{\text{GW}} &=&  2.5 \times 10^{-6} \left(\frac{100}{g_{*}(T_e)}\right)^{1/3} \Gamma^2 \bar{U}_f^4  
\left[
\frac{H_s}{\beta(v_w)}
\right]
v_w 
\times
\Upsilon 
% \nonumber \\ && 
\times S(f) , \label{eq:broken}
\end{eqnarray}
where $\bar{U}_f$ is the root mean square fluid velocity, $\Gamma \sim 4/3$ is the adiabatic index and ${\Upsilon}$ 
is the suppression factor arising from the finite lifetime~\cite{Ellis:2020awk,Guo:2020grp} ($\tau_{\text{sh}}$) of the sound waves~\cite{Guo:2020grp}
\begin{equation}
\Upsilon = 1 - \frac{1}{\sqrt{1 + 2 \tau_{\text{sh}} H_s}} .
\end{equation}
Finally, the spectral form has the shape
\begin{equation}
S(f) = \left( \frac{f}{f_p} \right)^3 \left( \frac{7}{4+3(f/f_p)^2} \right)^{7/2}
\end{equation}
where $f_p$ is the peak frequency given by 
\begin{equation}
    f_p = 8.9 \times 10^{-6} \frac{1}{v_w} \left(  \frac{\beta}{H_e} \right) \left( \frac{z_p}{10} \right) \left( \frac{T_e}{100 {\rm GeV}} \right) \left(  \frac{g_\ast (T_e)}{100} \right)^{1/6} {\rm Hz}
\end{equation}
However, a full calculation of the sound shell model can see qualitative deviations from this curve \cite{Hindmarsh:2019phv} with a better fit being a double broken power law. Most important for our interests is the fact that the power law after the peak depends on the strength of the phase transition and the bubble wall velocity \cite{Gowling:2021gcy}. A more optimistic scenario was studied in \cite{NANOGrav:2023hvm} where the power law on either side of the peak was treated as a free parameter. In this work, we will perform a full calculation of the sound shell model to take advantage of this flexibility in the peak of the spectrum. Note that in the sound shell model, by keeping the force term between the bubble wall and the plasma longer, the shape can be modified in the infrared~\cite{Cai:2023guc}. 

\begin{figure}[htbp]
  \centering
  \includegraphics[width=0.8\linewidth]{%nanogravfit.pdf
  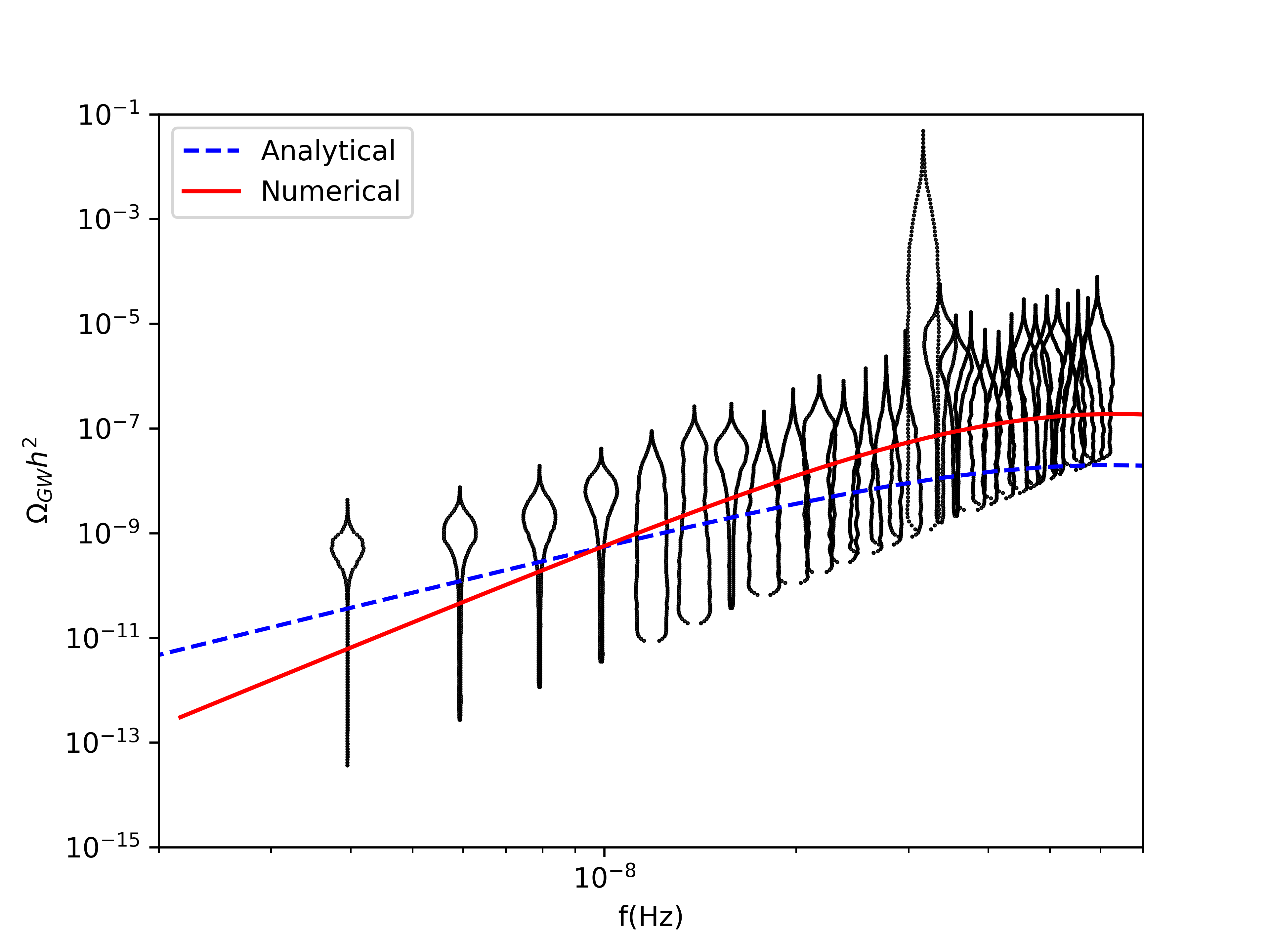
  }  \caption{\it The data from NANOGrav measurement for relic density of SGWB w.r.t. frequency in Hz (black) against the best fit using the full sound shell model (red) and the best fit for a broken power law fit (blue) frequently appearing in the literature (see eqn~\ref{eq:broken}). The parameters behind each fit are in Table~\ref{tab:parameters}.}
  \label{fig:fit-ng}
\end{figure}
\begin{figure}[htbp]
  \centering
  \includegraphics[width=0.8\linewidth]{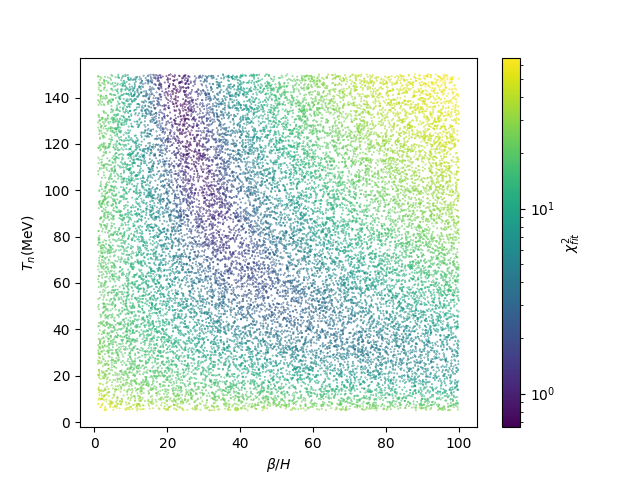}
\\
  \includegraphics[width=0.8\linewidth]{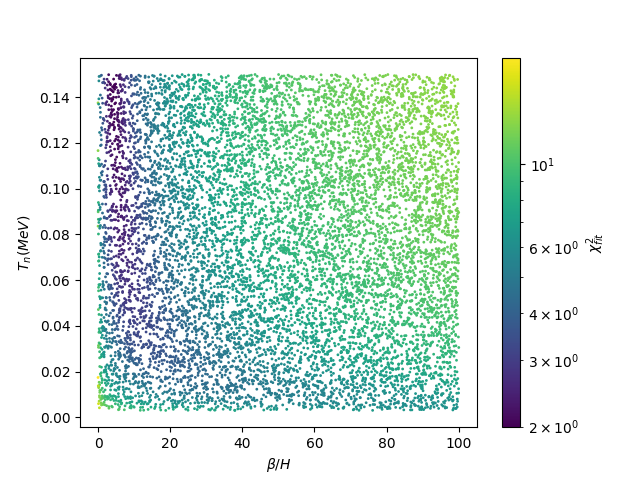}
  \caption{\it Scan over $(\beta /H_\ast, T_n)$ with $(\alpha , v_w)$ set to their best fit values for the full velocity profile (top panel) verse the broken power law fit (bottom panel). The fit favours problematic points with very small values for both parameters, with some parts of the preferred parameter space possibly threatening the fidelity of BBN, whereas the numerical calculation favours slightly larger values of $\beta/H_\ast$ and $T_n$ near the QCD transtion.}
  \label{fig:fit-ng2}
\end{figure}

\begin{table}[]
    \centering
    \begin{tabular}{c|c}
    \multicolumn{2}{c}{Full Sound shell} \\ \hline 
        Parameter & Best fit value  \\ \hline 
        $ \alpha _n$ & 0.85 \\
        $ \beta/H_\ast $ & 42 \\ 
        $ T_n $   & 133 {\rm MeV} \\ 
        $v_w$ & 0.09 \\ 
        $\chi _{\rm fit} $ & 1.4 \\
    \end{tabular}
        \begin{tabular}{c|c}
         \multicolumn{2}{c}{Broken power law fit} \\ \hline 
        Parameter & Best fit value  \\ \hline 
        $ \alpha _n$ & 0.89 \\
        $ \beta/H_\ast $ & 5.17 \\ 
        $ T_n $   & 142 {\rm MeV}\\ 
        $v_w$ & 0.67 \\ 
        $\chi_{\rm fit} $ & 1.59    
    \end{tabular}
    \caption{\it Best fit values for the full sound shell model and the usual fit used in the literature as given by Eq. \ref{eq:broken}. The full sound shell model performs somewhat better than the fit.}
    \label{tab:parameters}
\end{table}

\begin{figure}[htbp]
  \centering
  \includegraphics[width=0.8\linewidth]{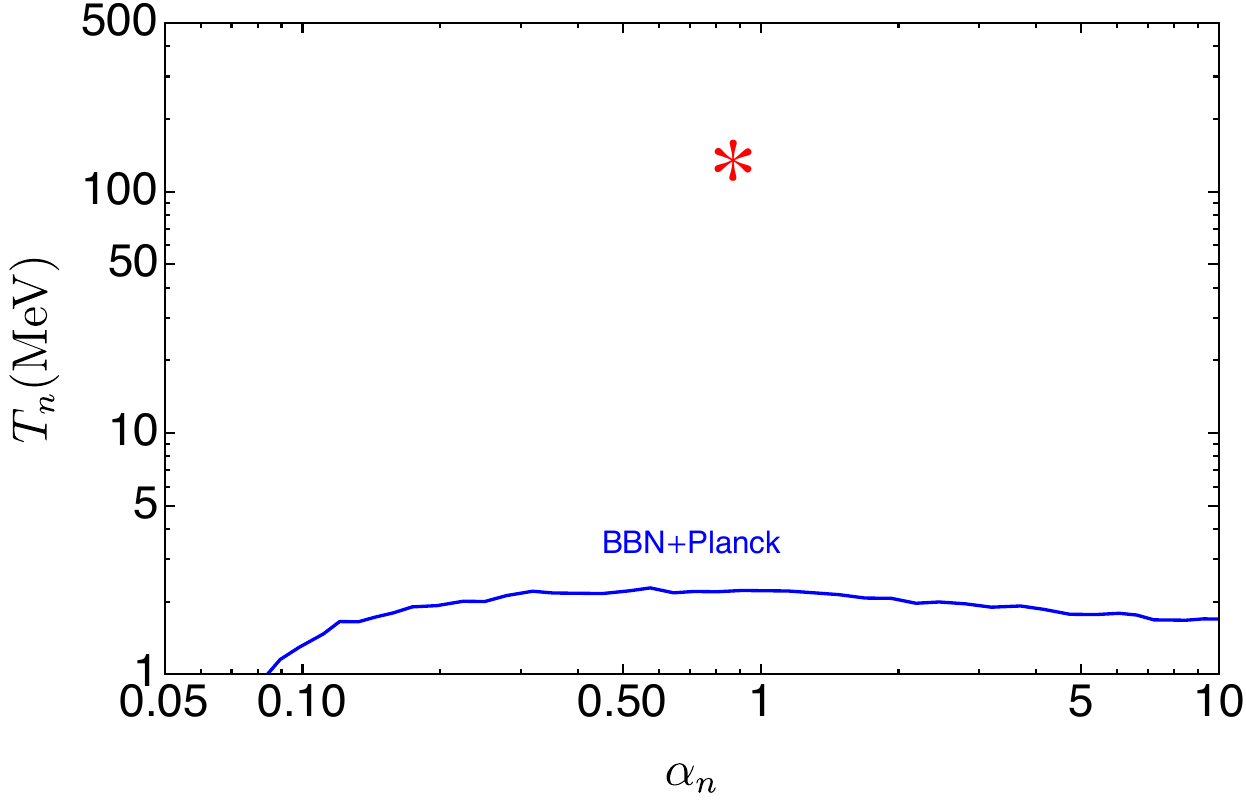}
  \caption{\it Constraints on FOPT parameters from PLANCK and BBN taken from Ref.~\cite{Bai:2021ibt}. The best-fit point obtained from our sound shell analysis with the NANOGrav data is shown by the red star.}
  \label{fig:Trh-bound}
\end{figure}

We perform a scan over the space of thermal parameters, $(\alpha _n, T _n, v_w, \beta/H_n)$, to find the best fit to the NANOGrav data (who have released their full data including uncertainties).  The scans  are  performed over the following ranges:  nucleation temperature $3~\text{MeV}<T_n<150~\text{MeV}$, bubble wall velocity $0<v_w<1$, phase transition strength $0<\alpha_n<1$, and  the efficiency of bubble formation w.r.t. the expansion rate $0<\beta/H<100$.  Since the relevant ranges of temperature and frequency are around the quark-gluon confinement regime near $150$~MeV,  we consider $g_*(T_n)$ the evolution of degrees of freedom for the energy density of the thermal bath of SM particles at the nucleation temperature  \cite{Drees:2015exa}. The best fit point we use the following figure of merit
\begin{eqnarray}
    \chi_{\rm fit}^2 = \sum_{i=1}^{N} \frac{\left(\log_{10} \Omega_{\rm th}h^2- \log_{10}\Omega_{\rm exp}h^2\right)^2}{2\bar{\sigma}_i^2}\,,
\end{eqnarray}
where $\Omega _{\rm th}h^2$ and $\Omega _{\rm exp}h^2$ represent the GW relic from theoretical prediction of FOPT and experimental value from PTA, respectively. Note that we ignore the width in the uncertainty regions, taking the midpoint and fitting to the vertical width. That is, $\sigma$, in the above equation is the distance from the midpoint value of $\log _{10}h^2 \Omega _{\rm GW}$ for each uncertainty region to the top or bottom. The best fit value of GW spectrum from our scan along with NANOGrav data are shown in figure \ref{fig:Trh-bound}.

To demonstrate the difference in preferred thermal parameters, in Fig.~\ref{fig:fit-ng2} 
we fix two thermal parameters to their best fit values and vary $\beta/H_\ast$ and $T_n$. The numerical scan prefers much more realistic values, slightly larger $\beta/H_\ast$ (it takes quite a fine tuned supercooling to go lower) and a temperature around the QCD transition. 
Using the data of NANOGrav we obtain the following values for the best fit point: $v_w\simeq0.09$, $\alpha_n\simeq0.85$, $T_n\simeq132.95$~MeV and $\beta/H\simeq42.02$. One note of caution, this is in the parameter range where there should be a large suppression due to energy lost to vorticity \cite{Cutting:2019zws}. In Appendix \ref{appendix1}, we show that this is somewhat mitigated by the fact that there are good fits which have a smaller trace anomaly and larger velocity.

\section{BSM Scenarios and Complementary Laboratory Probes}

We are somewhat spoilt for choice in models that can produce a strong first order phase transition at roughly the QCD scale. The very large strength of the transition lends credit to solitosynthesis as a possible explanation \cite{Croon:2019rqu}, as this mechanism typically leads to a stronger transition than conventional nucleation. The low wall velocity, however, supports a model that can predict a lot of friction like perhaps a SIMP model which can contain particles with large multiplicites \cite{Hochberg:2014dra,Hochberg:2015vrg,Garcia-Bellido:2021zgu,Chakrabarty:2022yzp}. Quite a few other dark sector phase transitions have been considered in this temperature range, see for instance ~\cite{Breitbach:2018ddu,Nakai:2020oit,Ratzinger:2020koh,Bai:2021ibt,Bringmann:2023opz,Deng:2023seh}. Of course, while the QCD phase transition is a crossover in the standard model at low density, a high lepton asymmetry or a different number of light quarks can change this picture ~\cite{Neronov:2020qrl,Li:2021qer,Sagunski:2023ynd}. We focus here on a dark sectors that have the prospect of having complementary probes in searches for long lived particles. A full model survey we leave for future work.

Let us now briefly discuss model-independent constraints on a MeV scale FOPT in the dark or hidden sector. During a FOPT, the vacuum energy contained in the false vacuum gets released, and a part of it  goes into reheating the photons or neutrinos in the plasma. The released energy may also end up heating relativistic particles in the dark sector. If the reheating of the SM particles happens at around or after the thermal decoupling  of neutrinos and photons, either or both of their temperatures will differ from the predictions of standard cosmology. This will change the relativistic degrees of freedom, $N_{\text{eff}}$, which is strongly constrained by Big Bang  Nucleosynthesis (BBN) and the Cosmic Microwave Background (CMB). The abundances of light element will also be modified and offer further bounds. $N_{\text{eff}}$ 
measurements severely constrain the dark sector reheating scenario as well. While our best fit point has a percolation temperature well above the scale at which we need to be concerned with BBN constraints, there are some points the agree well with NANOGrav data and have a much lower percolation temperature. There is  a distinction between percolation and nucleation temperatures \cite{Guo:2021qcq}. For a certain model, these temperatures can vary and produce different results. However, the current study uses a simple bag model for a physics scenario beyond the Standard Model, assuming a first-order phase transition where the nucleation temperature ($T_n$) is approximately equal to the percolation temperature ($T_p$). This assumption might change in other models, leading to different outcomes.

We first consider the reheating of the dark radiation case. In the approximation $T_{n}^D \ll T_{n}^{\gamma}$, one can show that $\alpha_n < 0.08$ for $T_{n}^{\gamma} \sim \mathcal{O}$(MeV)~\cite{Bai:2021ibt}. Ref.~\cite{Bai:2021ibt} also derived model-independent constraints on phase temperature ($T_n$) and strength parameter ($\alpha_n$) from $N_{\text{eff}}$ and helium and primary deuterium abundance ratios ($Y_P$ and $D/H|_P$, respectively) measured by CMB and BBN experiments when the FOPT heats the SM particles. For illustration, we discuss here the neutrino reheating scenario since the portal operator that can induce it (and the associated phenomenology) is relatively well-studied~\cite{Berryman:2022hds}. 
Ref. \cite{Deng:2023seh} explores the timing and temperature of the phase transition (PT) and its potential effects on BBN. Also, the nucleation rate, strength, and inverse duration of the PT are illustrated. The impact of the PT on the effective number of neutrino species ($N_{\rm eff}$) and neutrino decoupling is a critical concern, especially given the PT's temperature. However, the PT at temperatures well above the neutrino decoupling (over 100 MeV) does not adversely affect the neutrino-to-photon temperature ratio or increase $N_{\rm eff}$ beyond the acceptable cosmological limits. This is quantified by a function $F(t)$ in Ref. \cite{Deng:2023seh} that vanishes depending on the values $\beta/H$ and $T_n$. The vanishing of $F(t)$ can safely happen for our best fit points shown in Table \ref{tab:parameters}.

Using the BBN data from PDG~\cite{ParticleDataGroup:2020ssz} and the CMB data from the latest Planck results~\cite{Planck:2019nip},   Ref.~\cite{Bai:2021ibt} shows that the neutrino reheating temperature, $T^{\nu}_{\text{rh}}$, has to be greater than $\sim 3$ MeV for $\alpha_n > 0.1$. 
A future CMB experiment like CMB-S4~\cite{Abazajian:2019eic} will improve the bounds to $T^{\nu}_{\text{rh}} \gtrsim 4$ MeV. One can translate the above bounds on $T^{\nu}_{\text{rh}}$ to the phase transition temperature $T_{n}$ by using the formula,
\begin{equation}
    T^{\nu}_{\text{rh}} = \bigg[ 1 \, + \, \alpha_n \, + \, \alpha_n \frac{g_{*}^{\gamma}(t_{{\text{rh}}})}{g_{*}^{\nu}(t_{\text{rh}})} \Big( \frac{T_{n}^{\gamma}}{T_{n}^{\nu}} \Big)^4 \bigg]^{1/4} \, T_{n}^{\nu} \, \, ,
\end{equation}
where for reheating temperatures above MeV, $g_{*}^{\gamma}(t_{{\text{rh}}}) \approx 11/2$ and $g_{*}^{\nu}(t_{{\text{rh}}}) \approx 21/4$. Thus, one can conclude that the existing CMB and BBN data bounds place an almost flat constraint on $T_{n}^{\nu} \gtrsim 2$ MeV for $\alpha_n > 0.1$ as shown by the blue line in Fig.~\ref{fig:fit-ng}. The bound on $T_{n}$ from the photon reheating case is almost the same but extends to a bit smaller values of $\alpha_n$~\cite{Bai:2021ibt}. A cosmic phase transition at temperature $T_n$ can reheat photons and  neutrinos in different ways. The neutrino reheating will lead to equal reheating temperatures for photons and neutrinos, potentially resulting in a higher effective neutrino temperature due to the different degrees of freedom \cite{Bai:2021ibt}. To avoid any problem with the physics of BBN we will consider such a BSM scenario with phase transition.

\begin{figure}[htbp]
  \centering
  \includegraphics[width=0.8\linewidth]{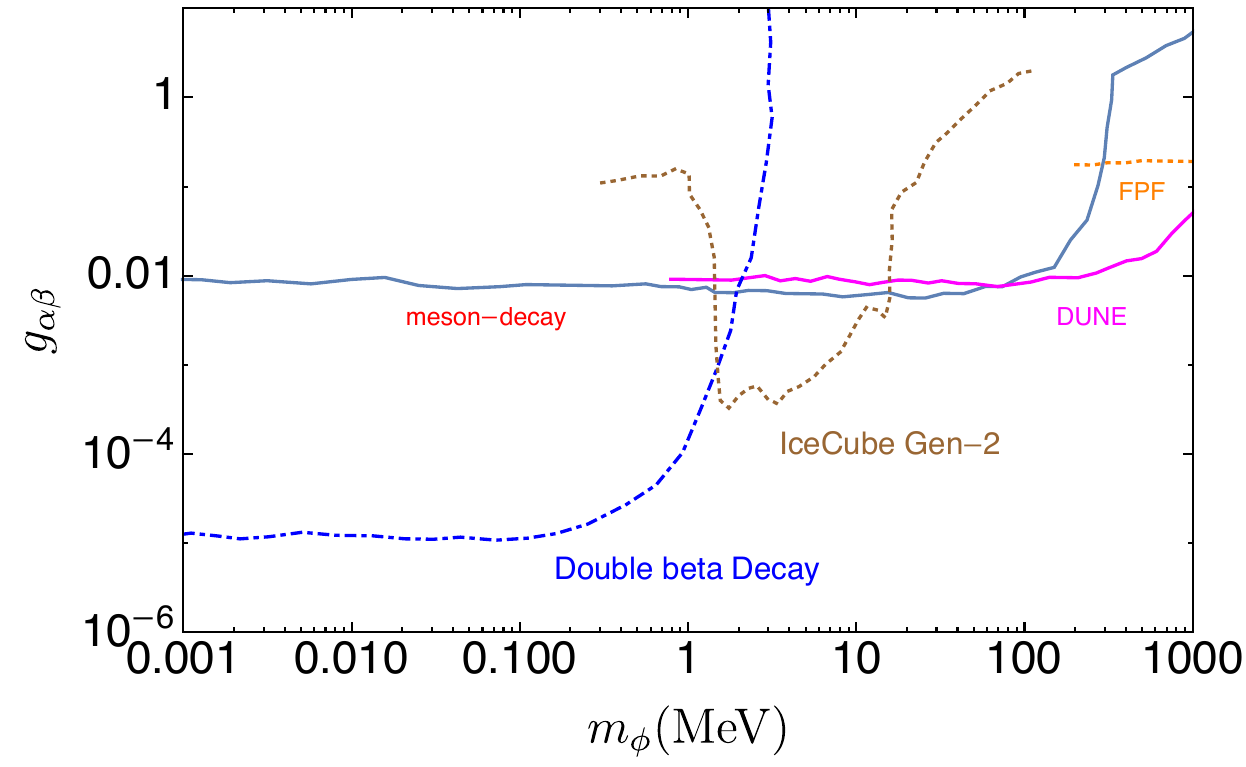}
  \caption{\it A selection of limits on the parameters $g_{\alpha \beta}$ and $m_{\phi}$ that can participate in reheating the neutrino plasma from the decay of a dark sector scalar. We show the existing laboratory constraints from meson decay spectra [dark blue solid] and neutrinoless double $\beta$-decay [blue dot dashed]. Also we show the projected limits from DUNE [magenta solid], IceCube Gen-2 [brown dotted] and forward physics facility at the LHC [orange dotted].  }
  \label{fig:fit-ng1}
\end{figure}

Finally, we provide a brief discussion on the interactions between the SM neutrinos and the dark sector scalar that is responsible for the FOPT under discussion. There is one point that we should clarify that the reheating has to be instantaneous for the above constraints to be applicable. For a delayed reheating, the constraints on the  FOPT is expected be much stronger~\cite{Bai:2021ibt}. The neutrino reheating can happen from the decay of a dark scalar $\phi$ to a pair of neutrinos via the dimension-6 effective operator
\begin{equation}
\label{eq:LSd6O}
    \mathcal{O} = \lambda_{\alpha \beta} \, \frac{(L^T_{\alpha} i \sigma_2 H) (H^T i \sigma_2 L_{\beta}) \phi}{\Lambda^2},  
\end{equation}
where $L$ and $H$ are the SM lepton and Higgs doublets, respectively. After the electroweak symmetry breaking the above operator will generate an interaction term $g_{\alpha \beta} \, {\nu} \nu \phi$, where $\alpha, \beta = e, \mu , \tau$  and $g_{\alpha \beta} = \lambda_{\alpha \beta} \, v^2/\Lambda^2$. Significant bounds already exists on the $g_{\alpha \beta}-m_{\phi}$ plane from existing laboratory experiments like meson decay spectra~\cite{Pasquini:2015fjv}, neutrinoless double $\beta$-decay~\cite{Brune:2018sab}, $Z$ or the SM Higgs invisible decay or $\tau$ decay~\cite{deGouvea:2019qaz}. Also, these couplings are significant interest of study in upcoming experiments like DUNE\cite{Berryman:2018ogk,Kelly:2021mcd}, generation-2 IceCube~\cite{Cherry:2014xra}, and forward physics facilities (FPF)~\cite{Kelly:2021mcd} at the LHC. We show a subset of these existing and projected bounds on the $g_{\alpha \beta}-m_{\phi}$ plane in Fig.~\ref{fig:fit-ng1}. For the detailed phenomenology of these couplings at various terrestrial and celestial experiments we refer the interested readers to the recent review paper on this topic~\cite{Berryman:2022hds}. As far as the UV-completion of the effective operator of Eq.~\ref{eq:LSd6O} is concerned, the most canonical models that can provide it are the massive Majoron models~\cite{Rothstein:1992rh,Gu:2010ys,Queiroz:2014yna,Brune:2018sab}. In addition, the generation of this effective operator from an inverse seesaw model~\cite{Lyu:2020lps} and a $U(1)_{B-L}$~\cite{Dev:2021axj} model has been considered in the literature.

\section{Discussions and Conclusions}

In this paper we have had an in depth look at sound wave induced gravitational waves from a strong first order cosmic phase transition as a possible explanation for the recent signal at multiple pulsar timing arrays. In particular, we have looked at how much including the full velocity profile rather than using a broken power law fit improves agreement with data. The best fit parameters also look a bit more realistic than what can be achieved via the broken power law, with the time scale of the phase transition being a smaller fraction of the Hubble time. We of course emphasize the caveat that understanding the spectrum from sound shell models is still in a state of flux. Reheating can suppress the nucelation rate enhancing the spectrum \cite{Jinno:2021ury}. On the other hand, energy lost to vorticity can suppress the spectrum \cite{Cutting:2019zws}. We leave a detailed analysis of this to future work. We then took a brief look at dark sector models that can be responsible for such a phase transition. We show that one for phase transitions occurring at low temperatures, the cosmological constraints from BBN, PLANCK data and future sensitivities from CMB experiments like CMB-S4, CMB-HD, CMB-Bharat, LiteBIRD will be complementary to the gravitational wave detectors to essential probe phase transition parameter space. This complementarity approach to probe phase transitions via GW detectors as well as CMB detectors paves the way distinguish the SMBHB and phase transition explanations to observed gravitational waves. Furthermore we showed that once we fix an operator that decides the interactions between the SM sector and the invisible sector (Eqn. 11) one is able to search for such mediators which is responsible for such interactions. We also discussed possible UV-complete neutrino mass models that can give rise to such low scale phase transitions and GW from sound waves measured in PTA data however detailed analysis involving a complete UV-complete model is beyond the scope of the current paper and will be taken up in a future publication. We envisage that the precision measurements that the GW cosmology and GW astronomy offers us from current data and from the planned worldwide network of GW detectors will make the dream of testing particle physics and fundamental BSM scenarios a reality in the very near future.

 Finally, we summarize the possibility of performing precision physics with NANOGrav. At the current level of precision of the data, there are several inputs in the phase transition process that can accommodate large variations. This is already evident from our Table 1, where we display the differences between a precise treatment of the sound shell model versus a fit to the broken power-law. Already at this level of precision, it is evident that there is a substantial difference in the benchmark values of the nucleation temperature, bubble wall velocity, and phase transition duration. In the future era of more data, a precise treatment of the PT will then have to take into account several quantities: $(1)$ source lifetime; $(2)$  mean bubble separation; $(3)$ going beyond the bag model approximation in solving the hydrodynamics equations and explicitly calculating the fraction of energy in the fluid from these equations rather than using a fit; $(4)$ including fits for the energy lost to vorticity modes; $(5)$ including fits for the energy lost to  reheating effects; $(6)$  a departure from the ``instantaneous reheating" approximation adopted in the study; a careful treatment of the difference between the  reheating and nucleation times, and the possibility of the nucleation temperature changing during the reheating process.

\medskip

\begin{acknowledgments}
% \paragraph{Acknowledgments}
The work of T.G. is supported by the funding available from the Department of Atomic Energy (DAE), Government of India for Harish-Chandra Research Institute. A.G. thanks hospitality of University of Pisa during the ongoing work. SFK acknowledges the STFC Consolidated Grant ST/L000296/1 and the European Union’s Horizon 2020 Research and Innovation programme under Marie Sklodowska-Curie grant agreement HIDDeN European ITN project (H2020-MSCA-ITN-2019//860881-HIDDeN).  KS is supported by the U.~S. Department of Energy grant DE-SC0009956. XW acknowledges the Royal Society as the funding source of the Newton International Fellowship. We thank Peter Athron for pointing out an error in the prefactor for the analytic fit that has propagated through the literature.
\end{acknowledgments}

% \clearpage
\appendix

% \onecolumngrid
\section{Scanning plots}
\label{appendix1}
In this section we provide more information about the preferred thermal parameter space in the numerical calculation of the soundshell model compared to the analytic fit that arises from taking the RMS fluid velocity. In particular, the wall velocity prefers to be large for the analytic fit to maximize the amplitude of the signal. However, as the shape depends upon the wall velocity and NANOGrav is best fitted to a reasonably gentle slope, the full numerical treatment prefers a relatively small wall velocity, $v_w \lesssim 0.2$.

\begin{figure}[htbp]
    \centering
    \includegraphics[width=0.48\linewidth]{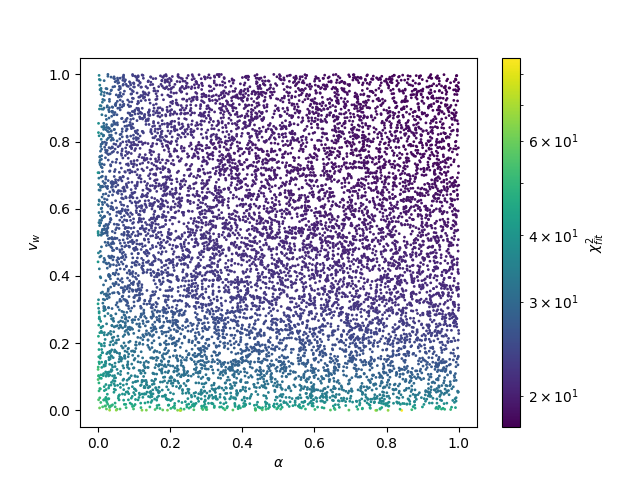}
    \includegraphics[width=0.48\linewidth]{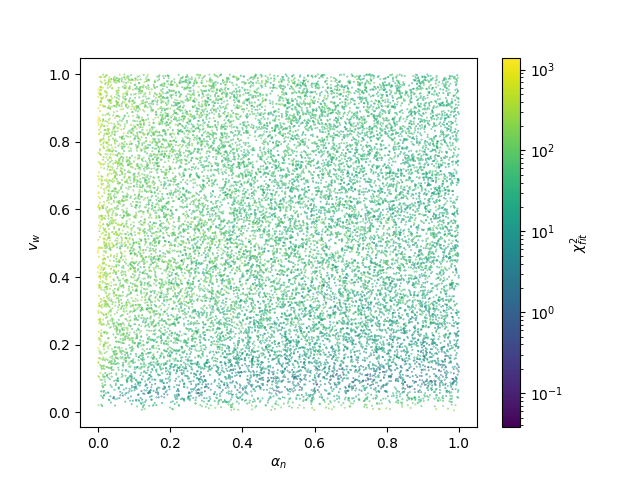}
    \caption{Scan over the thermal parameters $(\alpha , v_w)$ with $(\beta/H_\ast ,T_n)$ set to their best fit values for the broken power law fit (left panel) and the full numerical treatment of the soundshell model (right panel). Note that although the best fit point for the numerical treatment is in the danger region for the suppression factor from energy lost to vorticity modes to be large \cite{Cutting:2019zws}, there are many points with a small $\chi ^2$ that have a larger wall velocity and smaller $\alpha$.}
    \label{fig:vwalpha}

    \centering
    \includegraphics[width=0.48\linewidth]{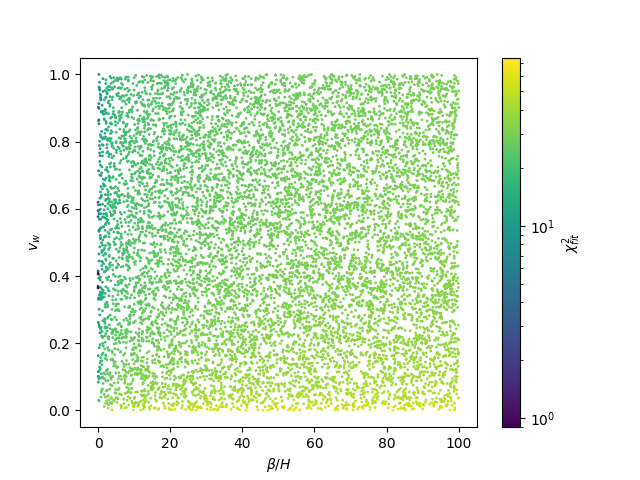}
    \includegraphics[width=0.48\linewidth]{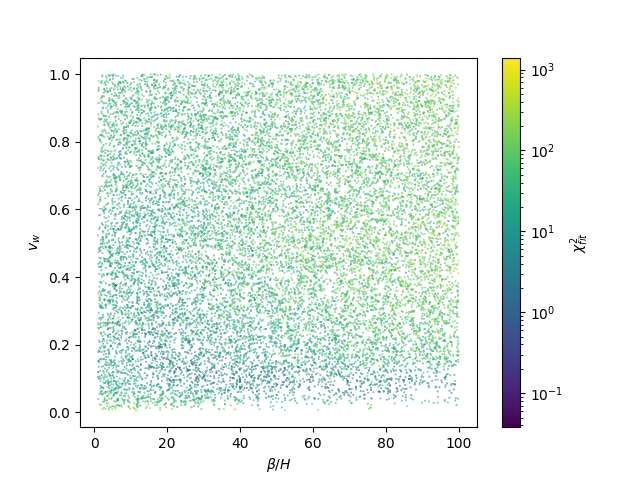}
    \caption{Scan over the thermal parameters $(\beta/H_\ast , v_w)$ with $(\alpha ,T_n)$ set to their best fit values for the broken power law fit (left panel) and the full numerical treatment of the soundshell model (right panel).}
    \label{fig:betavw}

    \centering
    \includegraphics[width=0.48\linewidth]{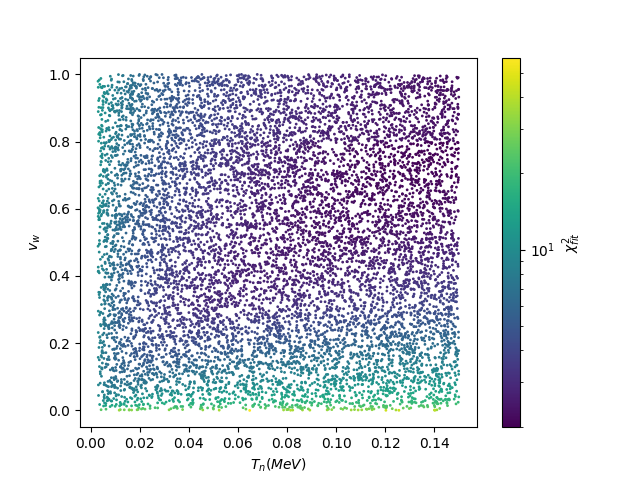}
    \includegraphics[width=0.48\linewidth]{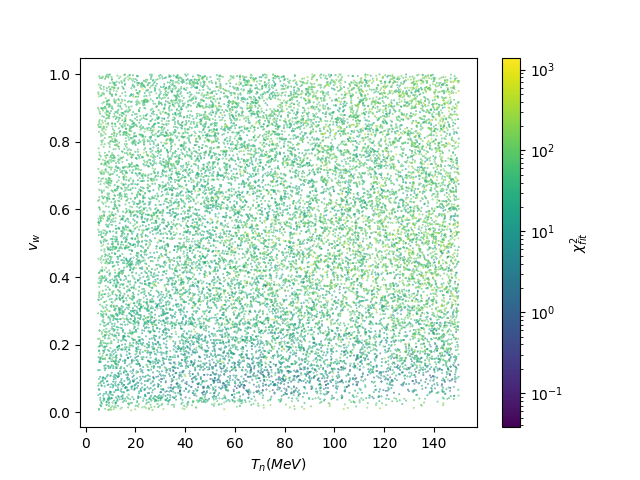}
    \caption{Scan over the thermal parameters $(T_n , v_w)$ with $(\beta/H_\ast ,\alpha)$ set to their best fit values for the broken power law fit (left panel) and the full numerical treatment of the soundshell model (right panel).}
    \label{fig:Tvw}
\end{figure}

One note of caution is the best fit values favour a low bubble wall velocity and at least a decent sized trace anomaly. Once the trace anomaly gets large, there is a substantial suppression for small wall velocities due to energy lost to vorticity modes \cite{Cutting:2019zws}. This is shown in figure \ref{fig:vwalpha}. This issue is mitigated somewhat by the fact that the $\chi ^2$ value remains smaller than the best fit value given by the broken power law even for small values of $\alpha$. Even still this motivates a detailed simulation to find the true preferred values of thermal parameters. Note that to avoid redundancy, we avoid repeating the plots depicting scans over $(\beta /H_\ast , T_n)$ which is included in the main body of the paper.

\begin{figure}[htbp]
    \centering
    \includegraphics[width=0.48\linewidth]{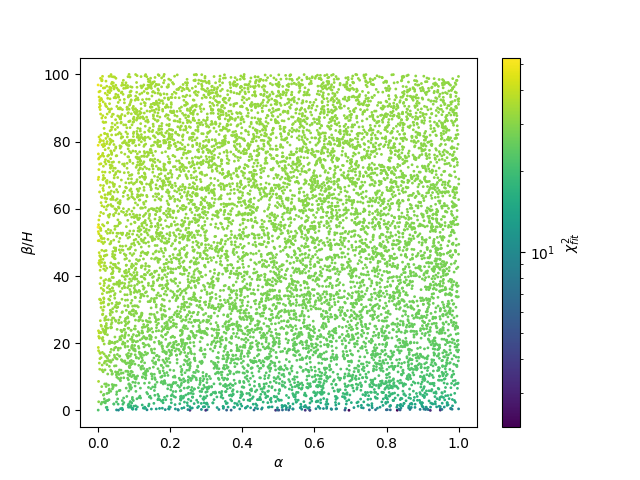}
    \includegraphics[width=0.48\linewidth]{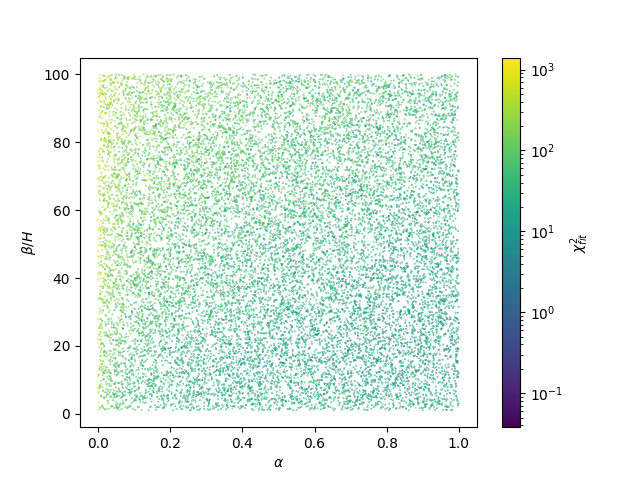} 
    \caption{Scan over the thermal parameters \( (\alpha , \beta/H_\ast) \) with \( (\alpha ,v_w) \) set to their best fit values for the broken power law fit (left panel) and the full numerical treatment of the soundshell model (right panel).}
    \label{fig:alphabeta}
\end{figure}
\begin{figure}[htbp]
    \centering
    \includegraphics[width=0.48\linewidth]{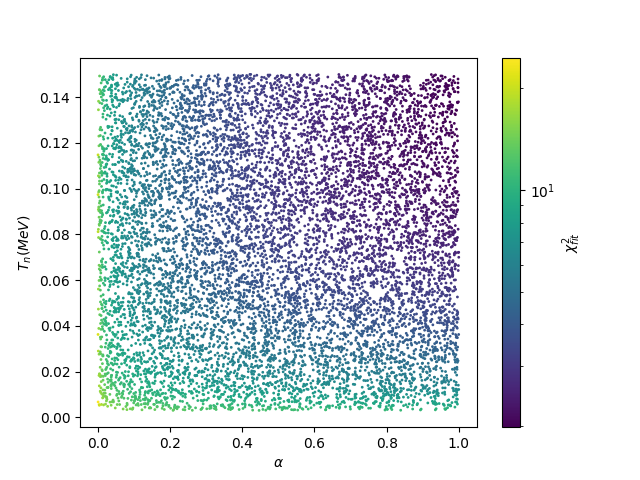}
    \includegraphics[width=0.48\linewidth]{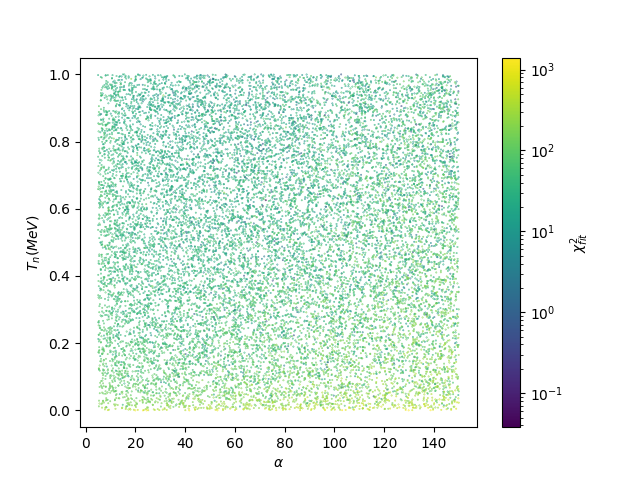}
    \caption{Scan over the thermal parameters \( (\alpha , T_n) \) with \( (\beta/H_\ast ,v_w) \) set to their best fit values for the broken power law fit (left panel) and the full numerical treatment of the soundshell model (right panel).}
    \label{fig:alphaT}
\end{figure}
%
%

%%%%%%%%%

\bibliographystyle{utphys}

\bibliography{theoretical, experimental, general}

\end{document}